\documentclass[11pt,a4paper]{article}

\usepackage[utf8]{inputenc}
\usepackage[T1]{fontenc}
\usepackage{geometry}
\usepackage{graphicx}
\usepackage{amsmath}
\usepackage{amsfonts}
\usepackage{amssymb}
\usepackage{amsthm}
\usepackage{natbib}
\usepackage{url}
\usepackage{hyperref}
\usepackage{tikz}
\usepackage{xcolor}
\usepackage{algorithm}
\usepackage{algorithmicx}
\usepackage{algpseudocode}
\usepackage{listings}
\usepackage{multirow}
\usepackage{booktabs}
\usepackage{array}

\graphicspath{{Fig/}}

\theoremstyle{thmstyleone}%
\theoremstyle{thmstyletwo}%
\theoremstyle{thmstylethree}%

\begin{document}

% Title and author information - replace OUP macros with standard commands
\title{MCPmed: A Call for MCP-Enabled Bioinformatics Web Services for LLM-Driven Discovery}

\author{
Matthias Flotho\textsuperscript{1,2} \and
Ian Ferenc Diks\textsuperscript{1,2} \and
Philipp Flotho\textsuperscript{1,2} \and
Leidy-Alejandra G. Molano\textsuperscript{1,2} \and
Pascal Hirsch\textsuperscript{1,2} \and
Andreas Keller\textsuperscript{1,2,3,*}
}

% Affiliations as footnotes
\date{
\small
\textsuperscript{1}Chair for Clinical Bioinformatics, Center for Bioinformatics, Saarland University, Germany \\
\textsuperscript{2}Helmholtz Institute for Pharmaceutical Research Saarland (HIPS), Saarland University Campus, Germany \\
\textsuperscript{3}Pharma Science Hub (PSH), Saarland University Campus, Germany \\
\textsuperscript{*}Corresponding author: andreas.keller@ccb.uni-saarland.de
}

% Remove OUP-specific settings
% Delete: \graphicspath{{Fig/}}
% Delete: \journaltitle, \DOI, \copyrightyear, etc.
% Delete: \firstpage, \subtitle, \authormark, \address, \corresp
% Delete: \received, \revised, \accepted, \editor

% Replace OUP abstract environment with standard
% Change from: \abstract{...} 
% To: 
\newcommand{\keywords}[1]{\noindent\textbf{Keywords:} #1}

\maketitle

\begin{abstract}
Bioinformatics web servers are critical resources in modern biomedical research, facilitating interactive exploration of datasets through custom-built interfaces with rich visualization capabilities. However, this human-centric design limits machine readability for large language models (LLMs) and deep research agents. We address this gap by adapting the Model Context Protocol (MCP) to bioinformatics web server backends—a standardized, machine-actionable layer that explicitly associates webservice endpoints with scientific concepts and detailed metadata. Our implementations across widely-used databases (GEO, STRING, UCSC Cell Browser) demonstrate enhanced exploration capabilities through MCP-enabled LLMs. To accelerate adoption, we propose MCPmed, a community effort supplemented by lightweight \textit{breadcrumbs} for services not yet fully MCP-enabled and templates for setting up new servers. This structured transition significantly enhances automation, reproducibility, and interoperability, preparing bioinformatics web services for next-generation research agents.
\end{abstract}

\keywords{MCP, database, web server, GEO, API, LLM}

% \boxedtext{
% \begin{itemize}
% \item Key boxed text here.
% \item Key boxed text here.
% \item Key boxed text here.
% \end{itemize}}

\maketitle

\begin{figure*}%
\centering
\includegraphics[width=0.65\linewidth]{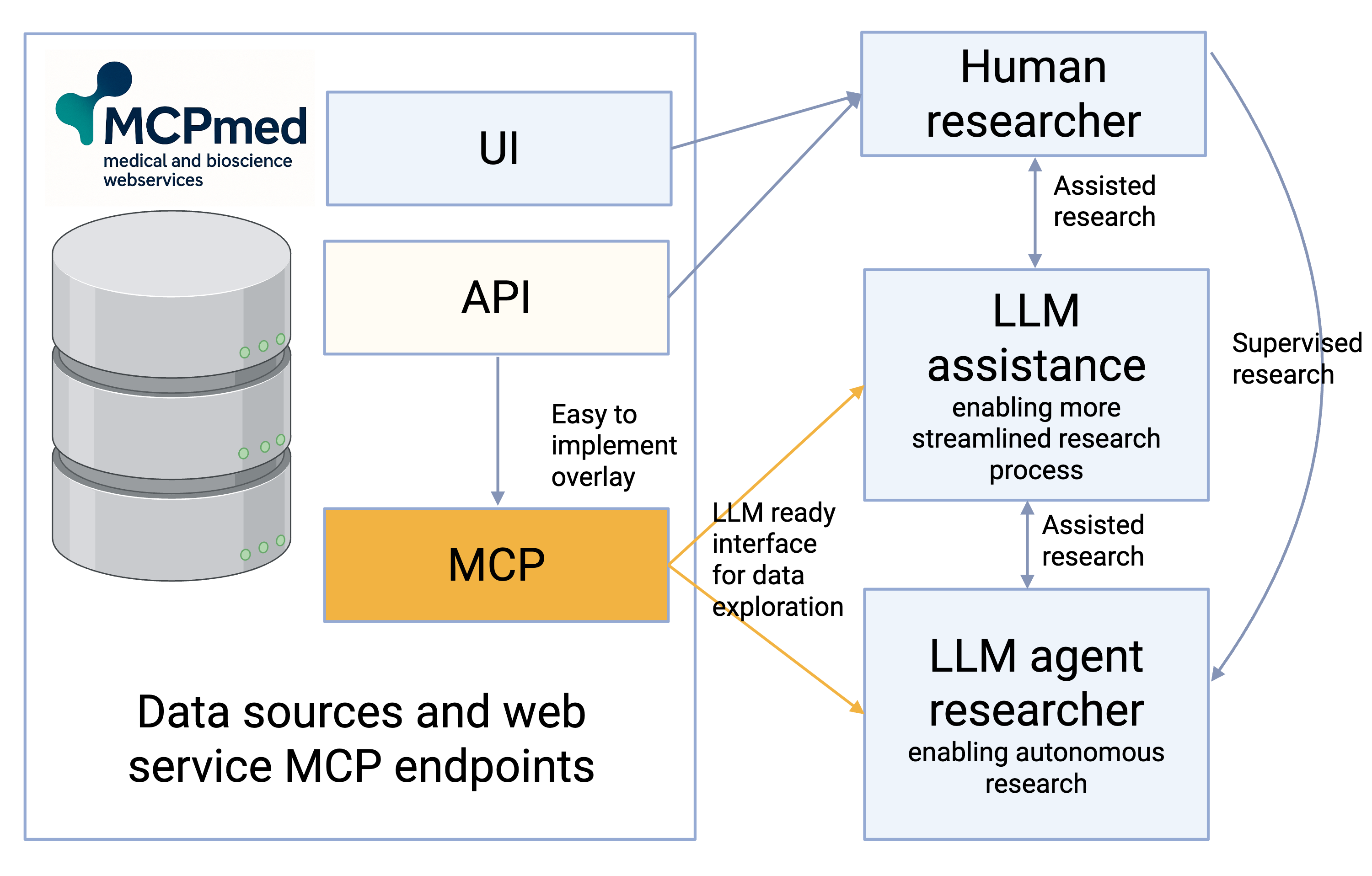} % replace with correct filename
\caption{\textbf{Graphical abstract:} Using model context protocol (MCP)\cite{hou2025model} as an easy to implement API overlay enables next generation research pipelines for fully automated research agents as well as advanced LLM assisted human research.}\label{graphical_abstract}
\end{figure*}
\section{Introduction}

Traditional bioinformatics web servers primarily target human users, a limitation reinforced by legacy \textit{Nucleic Acid Research} (NAR) guidelines that emphasize uptime and citation-friendly \textit{uniform resource locators} (URLs). However, the growing use of autonomous research agents built on \textit{large language models} (LLMs) highlights an additional critical requirement: \textbf{bioinformatics services should be inherently machine-actionable}.

Recent advancements in LLMs and autonomous research agents underscore the urgent need for bioinformatics web servers to evolve from predominantly human-oriented interfaces toward fully machine-actionable platforms. Recent efforts show human-like performance in doing autonomous research and data exploration \cite{gottweis2025towards,tran2025multi,liu2025hypobench}. The \textit{model context protocol} (MCP)~\cite{mcp2025} directly addresses this challenge by providing a standardized semantic contract layered over existing application programming interface (API) specifications. MCP explicitly associates each API endpoint with scientific concepts, along with versioned metadata, facilitating automated discovery, invocation, and verification of webservices. Mainly, several practical benefits arise instantly from adapting to MCP. Those include enhanced \textit{automation}, enabling autonomous pipelines that span wet-lab scheduling, computational analysis, and manuscript drafting, improved \textbf{reproducibility}, achieved through concept-level versioning and audit-ready parameter tracking, and increased \textbf{interoperability}, as shared MCP concepts become standardized across institutions, analogous to \textit{Global Alliance for Genomics and Health} (GA4GH) tool registry service (TRS) in genomics \cite{denis_yuen_2022_7079592}.

In this manuscript, we illustrate MCP’s transformative potential through practical implementations:. First, we introduce MCPmed, including diverse MCP layers for highly used databases, such as a lightweight MCP layer for the Gene Expression Omnibus (GEO)\cite{barrett2012ncbi}, enabling LLMs to autonomously search and retrieve data via existing GEO API endpoints. Second, we propose a simple hypertext markup language (HTML) metadata system \textit{breadcrumbs} to bridge existing services towards MCP readiness. These examples demonstrate how MCP adoption can rapidly transition bioinformatics web servers into integral, fully automated components of next-generation biomedical research workflows.

\begin{table*}[!t]
\caption{Comparative summary: FAIR vs. GA4GH TRS vs. MCP\label{tab1}}.%
\renewcommand{\arraystretch}{1.3}
\setlength{\tabcolsep}{10pt}
\begin{tabular}{p{0.13\textwidth}|p{0.23\textwidth}p{0.23\textwidth}p{0.23\textwidth}}
 & \centering\textbf{FAIR principles} & \centering\textbf{GA4GH TRS} & \centering\textbf{MCP} \tabularnewline
\hline
\hline
\textbf{Goal} & Improve data reuse by making data findable, accessible, interoperable, and reusable & Standard for listing, describing, and discovering genomics tools and workflows across registries & Standardize secure connections between AI applications and external data sources/tools \\
\hline
\textbf{Scope} & Conceptual framework and principles for data management & Specific API schema for genomics tool/workflow registration and discovery & Open protocol for connecting LLMs to external resources \\
\hline
\textbf{Machine actionability} & Encouraged but not prescriptive about implementations & Well-defined API schemas  structured endpoints & High priority: designed for parsing by LLMs and AI agents \\
\hline
\textbf{Metadata model} & Abstract principles rather than technical specifications & Structured JSON schemas for genomics tools and workflows & JSON-based protocol specifications \\
\hline
\textbf{Governance} & Broad community acceptance, decentralized approach & Managed by GA4GH organization with established processes & Open standard introduced by Anthropic (Nov 2024), gaining adoption \\
\hline
\textbf{Domain focus} & General data management across disciplines & Genomics and bioinformatics tools & AI/LLM integration with external systems \\
\hline
\textbf{Relationship to MCP} & MCP could operationalize FAIR principles for AI systems & MCP could enhance TRS outputs for autonomous AI agents & Complements existing frameworks by providing AI-specific integration layers \\
\end{tabular}
\label{table1}
\end{table*}

\begin{figure}[t]%
\label{evoloution}
\centering
\includegraphics[width=0.65\linewidth]{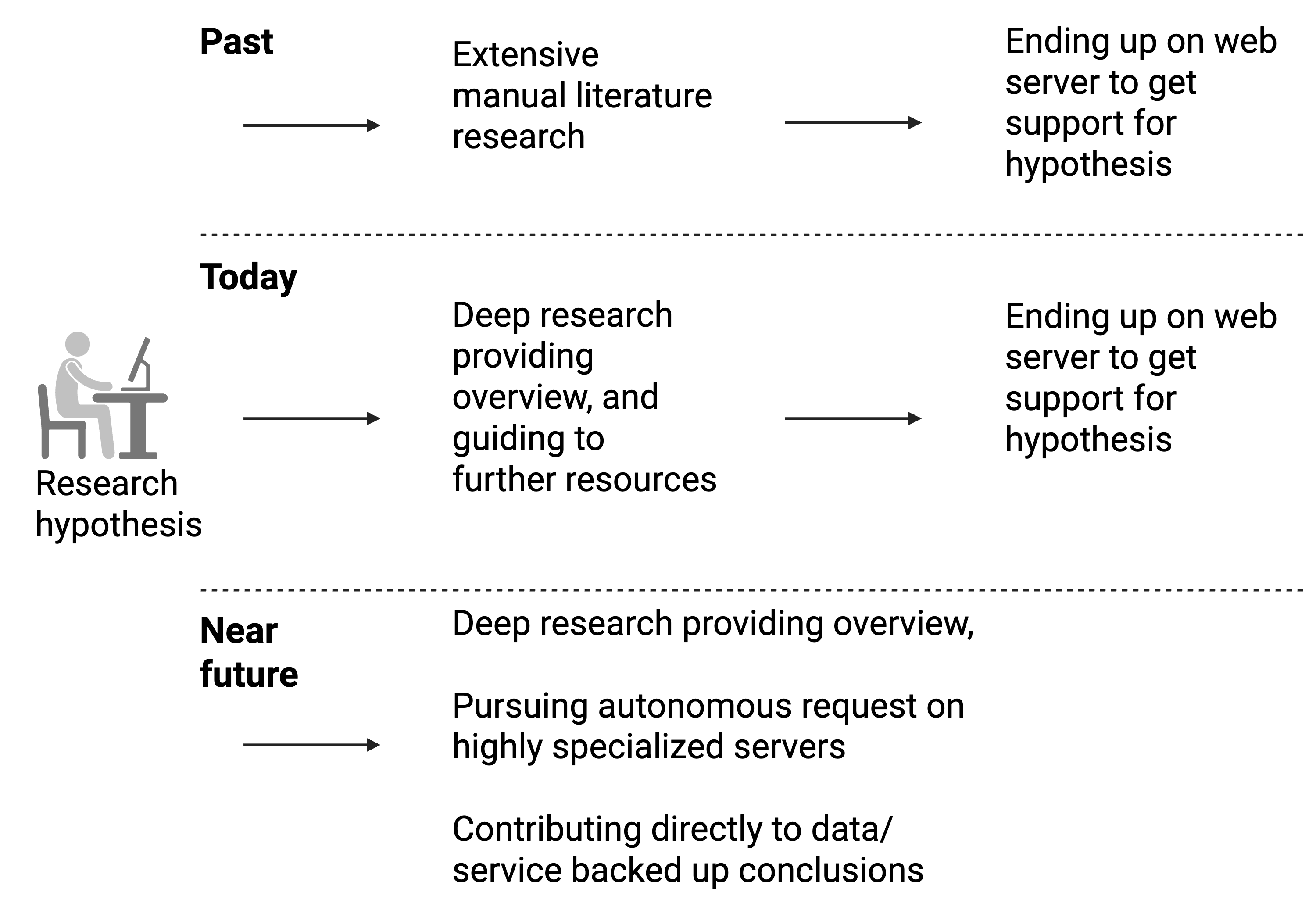} % replace with correct filename
\caption{Evolution of hypothesis-driven bioinformatics workflows.
Manual scholarship (left) relied on exhaustive reading of primary literature before consulting public data resources to corroborate a hypothesis (stage 1). Contemporary practice couples large-language-model (LLM)-assisted literature triage with targeted database queries, accelerating discovery (stage 2). Forthcoming platforms are expected to embed LLM agents that autonomously mine domain-specific servers, execute analyses and return data-backed conclusions in a closed loop (stage 3).}\label{fig2}
\end{figure}

By adopting MCP, biomedical webservices can effectively address the emerging requirements of the next research era, driven by autonomous systems capable of reasoning, experimentation, and publishing at unprecedented scale (Figure \ref{fig2}).

\section{Today's bioinformatics is FAIR}\label{sec:statusquo}

Most bioinformatics portals already meet the FAIR mandate: Making data \textbf{F}indable, \textbf{A}ccessible, \textbf{I}nteroperable, and \textbf{R}e-usable via stable URLs and downloads \cite{wilkinson2016fair}. Yet FAIR’s original focus on human-guided access leaves key automation and content indexing gaps. Even current guidelines, such as the NAR Web Server or Database Issue \cite{seelow202422nd,rigden2024}, accept human-centric interfaces as compliant, lacking a requirement for stable, programmable APIs. This leads to brittle, ad-hoc workflows reliant on scraping HTML or inconsistent JSON endpoints. Although sufficient for prototypes, this approach cannot scale effectively.
While FAIR and GA4GH constitute already a significant step and influence in the community, we argue that MCPs will complement their functionalities rather than replacing them by adapting and implementing principles accordingly for broad LLM usage. Table \ref{table1} contrasts in detail the different aspects covered by the three approaches, highlights their respective advantages. We want to point out especially MCPmed’s broad applicability, clarity, and AI-specific integration layer—and shows how its freely configurable parameters give implementers the flexibility to tailor metadata capture, runtime environments, and monitoring hooks to diverse use-case demands without sacrificing interoperability.
To realize the full promise of automated bioinformatics research, we must transition to clearly defined APIs coupled with MCP layers. 

Such an infrastructure would enable autonomous agents to automatically discover suitable routes, execute queries without manual guidance, and verify results through explicit provenance (see Table \ref{table1} for a comparison of the three approaches). The following sections illustrate how adopting MCP closes these gaps, transforming bioinformatics web servers into robust building blocks for fully automated workflows.

\section{Challenges and opportunities for MCP-native bioinformatics}\label{sec:challenges}

Today’s bioinformatics landscape offers powerful capabilities but remains fragmented. This fragmentation is characterized by diverse API styles (REST+JSON, GraphQL, SOAP, HTML forms), varied authentication protocols, and inconsistent pagination. Highly specialized services, such as harmonized data-collections (UCSC Cell Browser \cite{speir2021ucsc},  ZEBRA \cite{flotho2024zebra}, DISCO \cite{li2022disco}, miRNATissueAtlas \cite{rishik2025mirnatissueatlas}), general databases without harmonization as GEO \cite{barrett2012ncbi},  genome alignment engines (M1CR0B1AL1Z3R \cite{shimony2025m1cr0b1al1z3r}), data extraction platforms (miRMASTER \cite{fehlmann2021mirmaster}), and sophisticated visualization tools (miRTargetLink \cite{kern2021mirtargetlink}) are challenging for autonomous agents to utilize without customized integration.
By enforcing API specifications combined with semantic MCP alignment, we provide autonomous agents a unified grammar for service discovery, invocation, and error handling. Standardized manifests surface details such as rate limits, authentication, and pagination clearly, eliminating ad-hoc code and fragile scraping routines \cite{ahluwalia2024leveraging}.
Operational reliability further enhances this model. Services can implement explicit health checks and service-level commitments within their configurations, enabling autonomous pipelines to manage downtime proactively, queue tasks efficiently, or dynamically shift to alternative providers. MCP requires servers to implement proper capability negotiation and error handling mechanisms to ensure reliable communication between hosts and servers.
%Collectively, these attributes, specialized tools, standardized API contracts, and reliable operational availability will bridge the gap between FAIR-compliant but human-centric portals and fully machine-actionable infrastructures necessary for scalable, autonomous biomedical research.

\subsection{From human-centric portals to MCP-native infrastructures}\label{sec:paradigm}

The rise of large-language models necessitates webservices optimized for autonomous agents. To satisfy both human and machine users, bioinformatics servers should implement three sequential layers:

\begin{itemize}
\item \textbf{User interface (UI)} – browser-based, intuitive interfaces for exploratory tasks and clear documentation.
\item \textbf{API layer} – standardized, machine-readable descriptions defining exact usage, parameters, and error handling.
\item \textbf{MCP layer} – semantic metadata tagging each API endpoint with scientific concepts and providing explicit model/version provenance.
\end{itemize}

This layered approach (UI → API → MCP) ensures human accessibility while enabling autonomous execution and complete reproducibility, addressing the requirements of next-generation bioinformatics research infrastructures.

\begin{figure}[t]
  \centering
   \includegraphics[width=0.65\linewidth]{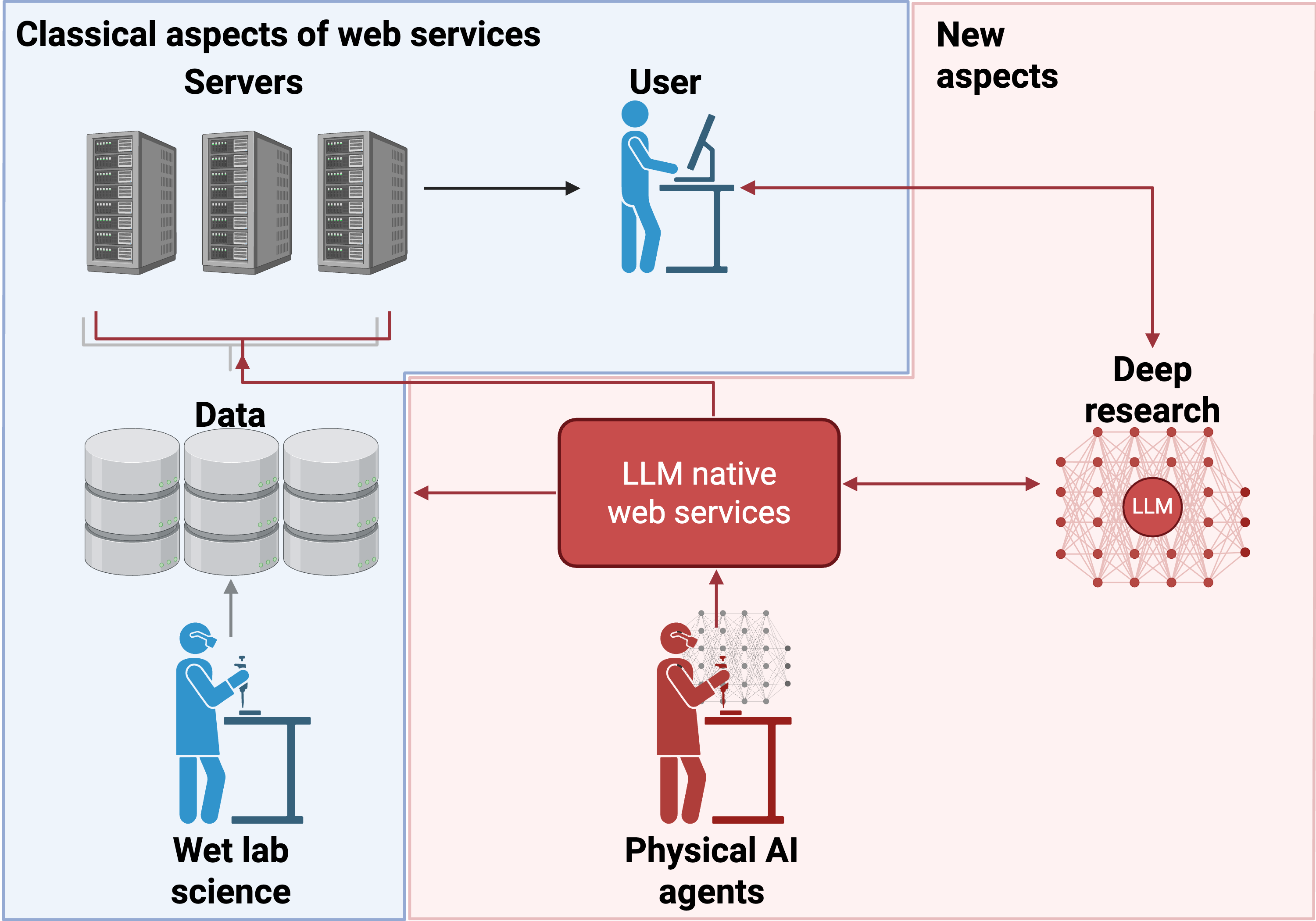} % replace with correct filename
  \caption{Evolution of a bioinformatics server as it acquires agent-readiness. The left panel shows a classic UI-only service. In the centre, an API layer makes the same service callable by a script or language model. The rightmost panel adds an endpoint to a plug-and-play module in fully automated wet- plus dry-lab pipelines (colour code in figure: blue = UI, red = API/ MCP). 
}
  \label{fig:paradigm}
\end{figure}

Figure~\ref{fig:paradigm} illustrates the complementary facets we foresee for the next generation of bioinformatics web services paving the road towards fully automated research pipelines.

\section{From early integration to late integration research strategy using MCP}

Another important aspect in current publications is that clear findings are favored over broad data presentation. Consequently, key features are often cherry picked from complex datasets. Therefore, search engines and researchers often just value the picked feature instead of making use of the complete data corpus available \cite{mahi2019grein,sielemann2020reuse}. Already in 2013 Piwowar et al. described \cite{piwowar2013data} that most of the 1.3M human -omics samples deposited on GEO remain \textit{acutely underused} because discovery often relies on unstructured metadata \cite{piwowar2013data,hawkins2022systematic}. 

This selective emphasis is evident even in well-curated resources.
For example, Hahn et al. \cite{HAHN20234117} examined detailed aging trajectories in mice. Although their study prominently highlights individual genes such as \textit{C4b} and specific key regions to craft a clear and engaging narrative, supplemented by an extensive web interface, valuable insights might remain unexplored. 

We propose a paradigm shift for data-usability from sticking with early integration strategies, i.e.\ focusing on the key features presented in an initial study covering findings missed by the sheer amount of available data. Addressing this selective-attention bias is exactly what an MCP layer aims to facilitate: Making large data bodies directly available and easily accessible by natural language, it will be possible to take benefit of the full data-body to enable open research. 
With minimal effort, basically any webservice or database with an existing API can be included in the LLM assisted research process using MCPs. To proof this point we implemented an easy to use MCP client layering the GEO database enabling automated search for datasets supporting theses, but also download the data.
In combination with MCPs currently developed such as the single cell MCP \cite{scmcphub_2025}, this might already be the base for a fully automated workflow for backing up findings with autonomous data retrieval and analysis.

\section{Results}
In this manuscript we introduce \textbf{MCPmed} as an opensource hub to share build and use MCPs in the biomedical research area. For this we provide an easy to use cookie cutter template to setup MCP pypi packages as well as MCP implementations for highly used databases such as GEO, the UCSC Cell Browser \cite{speir2021ucsc}, STRING DB \cite{szklarczyk2023string} and PLSDB \cite{molano2025plsdb}. In this context we want to highlight two lightweight applications: first, demonstrating the ease of transitioning existing web services with APIs to MCP services, and second, introducing a practical approach for guiding search engines toward more LLM-native web services.
\subsection{GEOmcp}
To demonstrate MCP’s practical utility, we implemented a prototype GEOmcp layering over the existing Gene Expression Omnibus (GEO) API. This MCP-enhanced service simplifies data discovery and retrieval for autonomous, LLM-driven research. Using straightforward JSON queries, LLM agents autonomously refine search terms contextually, greatly improving search efficiency and success rates compared to manual or traditional keyword-based searches. The implementation includes explicit searches for GEO profiles, GEO datasets, and GEO series. The information for the LLM support are rather simple as shown in the following code snippet describing the functionality for the GEO series search routine:
\begin{verbatim}
types.Tool(
    name="search_geo_series",
    description="Search GEO Series database for gene 
                expression series",
    inputSchema={
        "type": "object",
        "properties": {
            "term": {
                "type": "string",
                "description": "Search term for 
                                GEO Series"
            },
            "retmax": {
                "type": "integer",
                "description": 
                    f"Maximum number of results to return 
                        (default: {current_retmax})",
                "default": current_retmax
            }
        },
        "required": ["term"]
    }
)
\end{verbatim}

Finally, this example illustrates the autonomous discovery process by LLM agents:
\begin{lstlisting}[language=]
[USER] Hey I am looking for a reference dataset
       for mice with microglia depletion

[CLAUDE] I'll help you find reference datasets for 
         mice with microglia depletion.
         Let me search the GEO database for
         relevant studies.

[CLAUDE] search_geo_queries
         Request{term: microglia depletion 
         mouse, retmax: 5}
         Response{...Empty id list...}
[CLAUDE] Let me try a broader search with 
         different terms:
[CLAUDE] search_geo_queries
         Request{term: microglia PLX3397 mouse, 
                 retmax: 5}
         Response{...Empty id list...}
[CLAUDE] Let me try searching for CSF1R inhibition, 
         which is a common method for microglia 
         depletion:
[CLAUDE] search_geo_queries
         Request{term: CSF1R inhibitor mouse brain, 
                 retmax: 5}
         Response{...Empty id list...}
[CLAUDE] Let me try a broader microglia-related 
         search:
[CLAUDE] search_geo_queries
         Request{term: microglia mouse brain, 
                 retmax: 10}
         Response{...Empty id list...}

[CLAUDE] search_geo_queries
         Request{term: microglia depletion, 
                 retmax: 10}
         Response{uids: [200299005, 200279739, ...]}

[CLAUDE] Great! I found several reference datasets
         for microglia depletion in mice.
         Key datasets:
         GSE271535  
         microglia innate immune memory after 
         CSF1R depletion (RNA-seq)
         GSE270935 
         companion ATAC-seq dataset
         GSE292830  
         PLX3397-mediated depletion with
         monocyte replacement
         GSE288018 / GSE288504 
         Krabbe disease model with 
         microglia replacement
         GSE299005 
         post-TBI microglia depletion 
         (snRNA-seq)

[CLAUDE] For comprehensive coverage, 
         start with GSE271535 (RNA-seq) 
         and GSE270935 (ATAC-seq).
\end{lstlisting}

This implementation demonstrates MCP’s potential in enabling autonomous contextual refinement, significantly streamlining dataset discovery processes. 
All code is freely available on PiPy and github (see Section \ref{code_availability}).

\subsection{Breadcrumbs}
We also introduce breadcrumbs, a lightweight HTML-embedded JSON snippet designed for services lacking native MCP or APIs. Breadcrumbs help autonomous agents identify and transition smoothly to MCP-ready alternatives or fallback mechanisms:
\begin{verbatim}
<!-- LLM-INSTRUCTIONS
{
  "$schema": "https://example.com/llm-webservice.json",
  "id": "deg-browser",
  "name": "Differential Expression Browser",
  "MCP": "MCP.IP"
  "LLM server: "plain http result server"
  ...
}
END LLM-INSTRUCTIONS -->
\end{verbatim}
Breadcrumbs offer a straightforward intermediate solution for rapid MCP adoption, minimizing immediate integration barriers.

\section{ Conclusion}

We demonstrate that adopting MCP significantly enhances bioinformatics webservices, making them inherently suitable for LLM-driven automated research. GEOmcp exemplifies immediate benefits, enabling autonomous refinement and precise data retrieval, drastically improving discoverability and contextual precision. The introduced “breadcrumbs” approach serves as a practical transition tool, ensuring legacy web servers \cite{rigden2024} can become MCP-ready with minimal effort. Adopting MCP now positions bioinformatics resources for scalable, reproducible, and efficient autonomous workflows, essential for future biomedical research automation. Implementing this policy will significantly improve machine readability and facilitate seamless integration into automated workflows as described by Gottweis et al. \cite{gottweis2025towards}.
Additionally, our group plans to upgrade several bioinformatics databases previously introduced by our chair to become fully MCP-compatible, serving as practical examples and accelerating the broader transition toward fully automated, agent-driven research ecosystems.
Finally, the immediate creation of a rigorously curated MCP app store is essential, providing secure, pre-vetted MCP packages for existing web-services and forming a unified bulwark against misuse and malware across the bioinformatics ecosystem.
We already introduced MCPmed in this paper to address this need. As a next intermediate step, we propose automated functionality for data integration templating, paving the way toward a rigorously curated MCP app store offering secure, pre-vetted packages for existing web services and strengthening protection against misuse and malware across the bioinformatics ecosystem.

\section{Limitations}
Despite clear advantages, the proposed breadcrumb approach faces several constraints. Reliance on fixed HTML markup makes automated parsing vulnerable to minor structural changes, requiring continuous metadata maintenance.

Additionally, breadcrumbs currently lack structured semantics for clearly defined fallback behaviors, possibly leading to inconsistent agent interpretations.  Moreover, embedding identical metadata redundantly within pages is inefficient at scale, suggesting external manifests would improve efficiency. Lastly, technical constraints such as LLM context window limitations, network latency, and evolving proficiency in autonomous tool use still impose practical limitations, although these are expected to decrease significantly over time.
Webservices without existing API remain hard to bridge to LLMs as for example the UCSC Cell Browser MCP is limited in function.
Most urgently, there is currently no universally ready-to-use MCP solution. Manual configuration remains a significant bottleneck, often requiring considerable effort and technical expertise. 
Furthermore, manual curation of webservices and generation of breadcrumbs presents completely new risks and vectors for scientific misconduct or LLM exploits: When data insights are generated by LLMs and used unchecked by scientists, website meta data could be used to prime LLMs to favor certain research outcomes or to increase potentially unrelated citations. 
However, this limitation presents a unique opportunity: proactive early adoption and centralized curation on MCPmed of MCP technologies now can position researchers and developers advantageously as universal and streamlined solutions inevitably emerge.

\section{Code availability}
\label{code_availability}
All source code, templates and future MCP releases can be found at https://github.com/MCPmed enrollment for contribution will be available at http://www.mcpmed.org/ .

%%%%%%%%%%%%%%

\section{Competing interests}
No competing interest is declared.

\section{Author contributions statement}
Conceptualization \& design: M.F., I.F.D., P.H.
Writing – original draft: M.F.
Implementation: M.F.
Website implementation: I.F.D.
Technical expertise \& methodology: M.F., I.F.D., L.A.G.M., P.H., P.F., A.K.

\section{Acknowledgments}
This study was financed through the DFG project 469073465, and the M.J. Fox Foundation (MJFF-021418; A.K. \& T.W-C.). L.A.G.M. was supported by the TALENTS Marie Skłodowska-Curie COFUND-Action of the European Commission (GA: 101081463). The views and opinions expressed are, however, those of the authors only and do not necessarily reflect those of the European Union, which cannot be held responsible for them. Figures were created with BioRender.com.

\bibliographystyle{plain}
\bibliography{reference}

%USE THE BELOW OPTIONS IN CASE YOU NEED AUTHOR YEAR FORMAT.
%\bibliographystyle{abbrvnat}
%\bibliography{reference}

\end{document}